\documentclass[12pt,aps,tightenlines,showpacs,nofootinbib]{revtex4}

\usepackage{amsmath}
\usepackage{amsfonts}
\usepackage{amssymb}

\begin{document}

\title{Dark energy from coexistence of phases}
\author{Ariel M\'{e}gevand \\
Institut de F\'{\i}sica d'Altes Energies, Universitat Aut\`{o}noma
de Barcelona,  08193 Bellaterra (Barcelona), Spain}

\begin{abstract}
We suggest that the current acceleration of the universe may be
explained by the vacuum energy of a hidden sector which is stuck
in a state of equilibrium between phases. The phases are
associated to a late-time first-order phase transition, where
phase coexistence originates at a temperature $T_c \sim 10^{-3}eV$
and lasts until temperature falls below $T\sim 10^{-4}eV$. During
phase coexistence, the energy density has an effective
cosmological constant component with the observed magnitude. This
scenario does not require supercooling and may arise naturally in
realistic models.
\end{abstract}


\maketitle

\section{Introduction}

Evidence for the acceleration of the expansion rate of the universe,
coming from Type Ia supernovae observations \cite{obs}, together
with large scale structure and CMB measurements \cite{spergel06,
pr03}, are compatible with a flat universe which is composed of 76\%
dark energy and 24\% dark matter, with an energy density of order
$(10^{-3}eV)^4$. These observations raise some naturalness problems,
which could be summarized in two questions: why is there a tiny,
though non-vanishing, cosmological constant? Why do we happen to
live in the epoch in which such constant energy density coincides
with that of non-relativistic matter?

Many approaches to solve these problems assume that the
cosmological constant actually vanishes in the true vacuum due to
some as yet unknown mechanism, the observed vacuum energy being
caused by a field which is away from the global minimum of its
potential. This is the case, e.g., of quintessence models
\cite{pr03}, where a field $\phi $ is slowly rolling towards its
minimum.  An interesting possibility is that $\phi $ is a pseudo
Goldstone boson (PGB) which arises through the spontaneous
breaking of a global symmetry at a high scale $f$, and acquires a
harmonic potential when this symmetry is explicitly broken at a
lower scale $M$. The energy scale $M$ of the potential is
therefore $\sim \rho_\Lambda ^{1/4}$ in order to account for the
cosmological constant (this scale may be associated to a neutrino
mass), whereas the mass of $ \phi $ is suppressed by the scale $f$
in order to explain the slow evolution of the field \cite{axions}.

An interesting alternative to quintessence consists of models in
which the field is stuck in a false vacuum minimum. In this case,
a splitting $\varepsilon$ between the false and true vacua
accounts for the cosmological constant; therefore, we must have
$\varepsilon\sim \rho_\Lambda $, whereas the scale $M$ of the
potential can be larger. For instance, in models in which $\phi$
is a PGB, the potential has degenerate minima separated by
barriers of height $\sim M$. The splitting $\varepsilon$ can then
be achieved through higher-order effects \cite{gc93,bs01}. Another
possibility is that the false vacuum arises from non-perturbative
effects \cite{y02}. In all these cases we will have $\varepsilon
\ll M^4$. Alternatively, the false vacuum energy scale may be
obtained from the confining scale of a hidden gauge theory
\cite{g00}, or from $TeV$ scale supersymmetry breaking which is
Planck-scale suppressed in a hidden sector \cite{ahkm00,chn04}. In
these cases, the false vacuum energy arises from the spontaneous
symmetry breaking at the low scale, so the splitting between vacua
has the same scale as the potential, $\varepsilon \sim M^4$.

A general problem of false vacuum models is the difficulty to
accomplish that the field ends up actually in the false vacuum. In
the cases in which $\varepsilon \ll M^4$, the false vacuum is
stable under quantum tunnelling, because the splitting between
minima is much smaller than the barrier which separates them. In
these models, the field drops to one of the minima  at a high
temperature $T\sim M$, when it starts to feel the potential.
However,  it rolls toward each of the minima with equal
probability, and domain walls separating vacua are formed.  Then,
the regions with the true vacuum grow due to the pressure
difference between vacua. As a consequence, either the false
vacuum ends up disappearing, or a system of domain walls persists.
Inflation is usually invoked to solve this problem, but this
approach has nontrivial constraints \cite{bs01}. On the other
hand, in models with a single scale $ M^4\sim \varepsilon $, one
expects that thermal effects will favor the false vacuum for
$T\gtrsim M$, thus setting the proper initial condition for $\phi
$. However, in this case the probability of transition to the true
vacuum by quantum tunneling or thermal activation may be too high
for the false vacuum to survive long enough.

Indeed, such a single-scale false-vacuum model will generally have
a first-order phase transition at a critical temperature
$T_{c}\sim \rho_\Lambda ^{1/4}\simeq 2\times 10^{-3}eV.$ Notice
that whenever the temperature is close to any scale $M$ at which
there is a phase transition (i.e., $T\sim T_c\sim M$), the
radiation and vacuum energy densities will be comparable. Indeed,
before the phase transition, at $T>T_{c}$, the system is in the
high-temperature phase with $\rho _{R} \sim T_c^4$ and the field
is stuck in the false vacuum with $\rho _{\Lambda }\sim M^4$. This
fact is interesting, because the three forms of energy (matter,
radiation, and $\Lambda$) happen to be comparable within a
relatively short range of temperature scales $\sim (10^{-4}-1)eV$
\cite{ahkm00}. The coincidence of matter and radiation in this
range can be easily explained by a dark-matter particle at the
electroweak scale, whose relic density is governed by the scale
$M_\textrm{EW}^2/M_\textrm{Pl}\sim 10^{-3}eV$. On the other hand,
the single-scale models mentioned above present a potential with
non-degenerate minima at the scale $M\sim 10^{-3}eV$ (which could
also be related to the ratio $M_\textrm{EW}^2/M_\textrm{Pl}$).
Thus, one would expect a triple coincidence between radiation,
matter, and $\Lambda$ at $T\sim 10^{-3}eV$. In fact, the
coincidence is split by dimensionless weak coupling factors $\sim
1/\pi\alpha^2\sim 10^3$ which enter the dark-matter relic density.
Taking into account this correction, the matter-radiation equality
naturally becomes $T_{eq}\sim 1eV$ \cite{ahkm00}. Then, within the
framework of these models the radiation-$\Lambda$ equality will
occur at $T\sim T_c \sim 10^{-3}eV$ and, if the vacuum energy
remains constant, the matter-$\Lambda$ equality will occur at
$T\sim 10^{-4}eV$, in agreement with observation.

For this to happen, however, notice that, although the critical
temperature of the phase transition is naturally $T_c\sim
10^{-3}eV$, the false vacuum should not decay before the temperature
has fallen at least to $T\sim 10^{-4}eV$. Immediately after the
transition the field will be in the stable minimum of the free
energy. In general, the high-$T$ value of this minimum may differ
from its zero-temperature value. However, this minimum now evolves
with temperature, and its vacuum energy is relaxed to zero (or to a
lower scale value associated with a subsequent phase transition).
Furthermore, it is important to notice that the phase transition
occurs in a {\em hidden sector}, i.e, a system of particles which do
not interact with those of the standard model except through
gravity. In the discussion above, we have been assuming implicitly
that the temperature of this system is the same as that of photons.
This is not necessarily the case, since these systems are not in
contact. Moreover, the energy density of relativistic particles in a
hidden sector is constrained by big bang nucleosynthesis to be less
than $0.3 \rho_{\nu_e}$, where $\rho_{\nu_e}$ is the density of a
single species of left-handed neutrino (see \cite{g00} and
references therein). Hence, the temperature $T$ of the hidden sector
{\em must be lower} than that of photons $T_{\gamma }$ (how much
lower, depends on the number of extra species).  Since $T_{\gamma
}\simeq 2\times 10^{-4}eV \sim \rho_\Lambda ^{1/4}/10$, there seem
to be only two possibilities for us to witness the false vacuum:
either 1) the hidden sector has a hierarchy $T_{c}<\rho_\Lambda
^{1/4}/10$, allowing its temperature $T$ to fulfill the double
constraint $T_c<T<T_\gamma$ \cite{chn04}, so that the phase
transition has not occurred yet, or 2) the critical temperature
satisfies the more natural relation $T_{c}\sim\rho_\Lambda ^{1/4}$
but the system is supercooled, which means that the nucleation of
bubbles has not begun yet, even though the temperature has fallen
well below $ T_c$. In the latter case the supercooling must be of at
least an order of magnitude, $T<T_\gamma\sim T_{c}/10$
\cite{g00,ch00}. Both possibilities turn out to be quite unnatural
and strongly constrain the model.

In this letter we wish to suggest an alternative scenario for
single-scale false-vacuum models. We will study the possibility that
the development of the phase transition is very slow, so it is not
complete yet, even though bubble nucleation  began when the
temperature of the hidden sector was $T\simeq T_c\sim \rho_\Lambda
^{1/4}$. In this way, there is no need to force a split between
$T_c$ and $\rho_\Lambda ^{1/4}$, nor to require an excessive amount
of supercooling. Since at $T=T_{c}$ both the high- and
low-temperature phases have non-vanishing potential energy, an
average dark-energy component will remain while the two phases
coexist. In the next section we describe the scenario, and in
section \ref{models} we compare it with the other alternatives,
within the framework of two definite models. Our conclusions are
summarized in section \ref{conclu}.

\section{The phase-coexistence scenario}

It was pointed out by Witten \cite{w84} that a first-order phase
transition may occur reversibly in the universe. The essential
idea is that the energy (latent heat) that is expelled by the
expanding bubbles of low-temperature phase may keep the two phases
in equilibrium at $T=T_{c}$ until the phase transition is
completed. In fact, there is always some supercooling, and the
phase transition begins when $T\lesssim T_{c}$; however, the
entropy that is released in a first-order phase transition can
reheat the system up to a temperature $T$ very close to $T_c$.
Then, the pressure difference between the two phases becomes very
small and the phase transition slows down significantly
\cite{m04}. If the phase transition occurs in a hidden sector,
this system will be kept in phase equilibrium at constant $T\simeq
T_{c}$ while the temperature of photons decreases.

It is important to remark that the temperature $T$ of the hidden
sector must be initially lower than the temperature $T_\gamma$ of
photons. Indeed, the hidden sector has a radiation component which
should never dominate the cosmic expansion law. At $T=T_c\sim
10^{-3}eV$, the energy density of this component is\footnote{Notice
that we call $T$ and $\rho_R$ respectively the temperature and
radiation density {\em of the hidden sector}. For the CMBR we use
instead $T_\gamma$ and $\rho_\gamma$.} $\rho_R \sim T_c^4\sim
\rho_\Lambda$. Notice however that, when this equality between
$\rho_\Lambda$ and $\rho_R$ occurs in the hidden sector, since
$T_\gamma
> T$,  the energy
density of photons is $\rho_\gamma \sim T_\gamma^{4}\gg \rho_R$.
Then, the phase transition begins, and enters the slow
phase-coexistence stage. A vacuum energy density $\sim
(10^{-3}eV)^4$ will remain until the end of the phase transition. In
the meantime, the CMB radiation density $\rho_\gamma$, and the
matter density, decrease. When $T_\gamma$ becomes $\sim 10^{-3}eV$,
the equality $\rho_\gamma = \rho_\Lambda$ occurs; finally, at
$T_\gamma\sim 10^{-4}eV$ the matter density becomes comparable to
$\rho_\Lambda$.

Notice that, even if at late times the species are not in thermal
equilibrium with each other, as long as they have thermal
distributions they contribute to the finite-temperature effective
potential. During the phase transition, there is a back-reaction on
the particle distributions. The released entropy keeps the kinetic
energy of particles high and prevents their temperature from
decreasing.

The low-temperature phase has a lower energy density than the
high-temperature one. The energy difference is the latent heat
$L$, which is liberated as bubbles of the low-temperature phase
expand. In general, bubble nucleation begins at a temperature
$T_{N}\lesssim T_{c}$. If $L$ is comparable to the density $\rho
_{R}$ of the relativistic gas in the hidden sector, the system
reheats up to a temperature very close to $T_c$ and remains so for
a long time \cite{m04}. No further bubbles nucleate; the two
phases coexist at a constant temperature $T\simeq T_{c}$ as
bubbles of low-temperature phase slowly expand at the expense of
the regions of high-temperature phase \cite{w84}. We will assume
for simplicity that supercooling and reheating happen in a
negligibly short time and consider only the subsequent
phase-equilibrium stage \cite{s82}, which is in general longer. We
will thus obtain a lower bound for the total duration of the phase
transition.

As we will see, the duration of phase coexistence depends
essentially on the amount of latent heat that is released, and on
the energy density of radiation. In the early universe, there is a
hot plasma with a large number $g_*$ of relativistic species. In
general, only a few, say $g$, of them contribute significantly to
the effective potential of a field which undergoes a phase
transition (e.g., those which have the strongest Yukawa couplings).
Thus, the latent heat is proportional to $g$, whereas the radiation
density is proportional to $g_*$. As a consequence, the entropy that
is liberated in the transition has a relatively small effect on the
plasma, due to the large heat capacity of the latter. On the
contrary, at late times and low temperatures, only a few species
remain relativistic. Furthermore, since the different species are
not in thermal equilibrium with each other, it is likely that the
released entropy goes only to the $g$ particle species that
contribute to the effective potential, and no entropy is transmitted
to the rest of the species. Consequently, the released heat will be
comparable to the density $\rho_R$ of this sector. Therefore, the
occurrence of phase coexistence is favored at later epochs.

\subsection{The phase transition}

We will thus assume that the true vacuum energy vanishes at zero
temperature, and consider a $meV$ scale effective potential with a
first-order phase transition in a hidden sector with $g$
relativistic degrees of freedom. The general scenario depends only
on thermodynamic parameters such as the energy density and the
latent heat, so for the moment we will not need to specify any
model. Later we will consider a couple of examples.

Before the phase transition, the energy density of the hidden
sector is of the form $\rho =\rho _{R}+\rho _{\Lambda }$, where
$\rho _{R}=g\pi ^{2}T^{4}/30$  corresponds to radiation, and $\rho
_{\Lambda }$ is the constant energy density associated to the
false vacuum. The pressure is thus given by $p=\rho _{R}/3-\rho
_{\Lambda }$. To determine the duration of phase coexistence, we
use the adiabatic expansion of the universe \cite{m04}. The
entropy density of the hidden sector is given by
\begin{equation}
s=s_{+}\left( a_{i}/a\right) ^{3},  \label{adiab}
\end{equation}%
where $a_{i}$ is the scale factor at the beginning of the phase
transition, and $s_{+}=4\rho _{R}\left( T_{c}\right) /3T_{c}$ is
the entropy density of the high-temperature phase at critical
temperature. Normally, the temperature would decrease as entropy
dilutes. During a first-order phase transition, however, energy is
released in the form of latent heat $L\equiv T_{c}\left(
s_{+}-s_{-}\right)$, thus avoiding the temperature decrease.
During phase coexistence, the average entropy density is given by
$s=s_{+}+\left( s_{-}-s_{+}\right) f$, where $f$ is the fraction
of volume that has already converted to the low-temperature phase.
It follows that
\begin{equation}
f=\frac{ s_+}{s_+-s_-} \left[ 1-\left( a_{i}/a\right) ^{3}\right]
. \label{frac}
\end{equation}%

The phase transition completes when $f=1$. Since $s_->0$, this
occurs for a finite $a=a_f$. However, in the limit $s_-\to 0$ the
duration of the phase transition becomes infinite, so $a_f$ can be
made arbitrarily large if most particles in the hidden sector lose
their entropy. As we shall see, such a strongly first-order phase
transition is possible if the particles acquire large masses and
become non-relativistic \cite{cmqw05}. The condition $s_-\ll s_+$
implies that  the latent heat must be close to its upper bound,
which corresponds to $s_-=0$,
\begin{equation}
L\simeq L_{\rm max}\equiv 4\rho _{R}/3. \label{lrho}
\end{equation}%
We have seen that the temperature of photons must have decreased
by at least an order of magnitude since the beginning of the phase
transition in the hidden sector, so we must require that the
transition does not complete before $a/a_{i}> 10$. We thus obtain
the condition  $s_-<10^{-3}s_+$.

In the above we have implicitly assumed that the latent heat is
spread out quickly enough to guarantee a uniform temperature at
any time. This will be the case if the time scale for bubble
growth is much longer than the time required for latent heat to
travel the distance between bubbles \cite{ma05}. This condition
can be expressed as
\begin{equation}
df/dt\ll c_{s}/d,  \label{homog}
\end{equation}%
where $c_{s} \approx 1/\sqrt{3}$ is the speed of sound in the
relativistic gas, and $d$ is the average bubble separation, which
is roughly determined by the number density of bubbles $n_b$,
$d\sim n_{b}^{-1/3}$. During phase equilibrium at $T\simeq T_{c}$,
the number of bubbles is fixed, so $d\propto a$ and the rhs of Eq.
(\ref{homog}) decreases like $a^{-1}$. On the other hand, from Eq.
(\ref{frac}), $df/dt\sim \left( \rho _{R}/L\right) \left(
a_{i}/a\right) ^{3}H$; so, the lhs of Eq. (\ref{homog}) decreases
more quickly than the rhs. Thus, this condition will remain valid
in the range of interest if it is fulfilled at the beginning of
the phase-equilibrium stage. For $a=a_{i}$, Eq. (\ref{homog})
becomes $\rho _{R}/L\ll n_{b}^{1/3} H^{-1}$. Notice that
$n_{b}H^{-3}$ is the number of bubbles nucleated inside a causal
volume. We certainly have $n_{b}H^{-3}>1$ if the phase transition
has begun (this is the standard condition for a rough estimation
of the onset of nucleation). Moreover, when the phase-coexistence
stage is reached all bubbles have already nucleated, and this
number is in general $\gg 1$ (see, e.g., Refs. \cite{m04,ma05}).
On the other hand, for $L\simeq 4 \rho _{R}/3$ we have $\rho
_{R}/L\lesssim 1$. Hence, the requirement (\ref{homog}) is
generally satisfied.

\subsection{The equation of state for phase coexistence}

During the phase transition, the temperature and pressure have
constant values $T\simeq T_{c}$, $p\simeq p_{c}\equiv p(T_c)$, so
the energy is given by $\rho =T_{c}s-p_{c}$, with $p_{c}=\rho
_{R}\left( T_{c}\right) /3-\rho_\Lambda$. Using again Eq.
(\ref{adiab}), we
obtain%
\begin{equation}
\rho =\,T_{c}s_{+}\left( a_{i}/a\right) ^{3}-p_{c}.
\label{density}
\end{equation}%
This result can also be obtained by considering the average $\rho
=\rho_- f+\rho_+(1-f)$, where $\rho_+ = \rho_R(T_c)+\rho_\Lambda$,
$\rho_-=\rho_+-T_c \Delta s$, and $f$ is given by  Eq.
(\ref{frac}). According to Eq. (\ref{density}), as far as the
Friedmann equation is concerned, this system can be thought of as
being composed of a pressureless fluid, whose density $\rho
_{M}^\textrm{eff}=(4\rho _{R}/3)\left( a_{i}/a\right) ^{3}$
dilutes like matter, plus a constant energy density $\rho
_{\Lambda }^\textrm{eff}=-p_{c}$. The effective cosmological
constant is thus given by
\begin{equation}
\rho_{\Lambda }^\textrm{eff}=\rho_\Lambda -g\pi ^{2}T_{c}^{4}/90.
\label{Leff}
\end{equation}%
If $T_c\sim\rho_\Lambda^{1/4}$, we will have $\rho_{\Lambda
}^\textrm{eff}>0$ provided that $g$ is not too large. Hence we have
a constant $\rho_{\Lambda }^\textrm{eff}\sim
\left(10^{-3}eV\right)^4$ throughout the phase transition.

\subsection{Possible signatures}

It is interesting that, although the system is composed only of
vacuum energy and radiation, the density $\rho$ of the hidden
sector exhibits a matter component. One may wonder whether it
could account for dark matter. Notice however that this effective
matter density is not clustered. Such a homogeneous
$\rho^\textrm{eff} _{M}$ could provide a signature of phase
coexistence. However, it is too small to be perceived in current
observations. Indeed, it was $\rho _{M}^\textrm{eff}\sim \rho
_{R}\sim \rho _{\Lambda }$ at $a=a_{i}$. But at that stage the
{\em total} dark matter density of the universe was
$\rho^{\textrm{tot}}_{M}\sim (a_0/a_i)^3\rho _{\Lambda }$. Since
$a_0/a_i$ is at least $\sim 10$, we have $\rho^\textrm{eff}
_{M}<10^{-3}\rho^{\textrm{tot}}_M$.

Although the effective $\omega(z)$ in this picture is not
observably different from constant $\omega=-1$, our scenario is
characterized by the coexistence of two phases with a constant
${\cal O}(1)$ difference in the value of the vacuum energy. Such
kind of inhomogeneity in the cosmological constant will result in
specific signatures in large scale structure, which may allow a
distinction from $\Lambda$CDM and from quintessence. Since these
inhomogeneities become important at $z\sim 1$, they will not
influence structure formation on scales below that of galaxy
clusters. If their size is on the cluster scale or above, they may
have an effect on cluster abundance, causing a departure from the
results of a constant-$\Lambda$ model. Nevertheless, we expect our
scenario to cause deviations no larger than those of models in
which dark energy is coupled to matter and clusters in overdense
regions. Current data on cluster number counts do not allow to
discriminate dark energy models.

Further signatures might be found in the CMB, which may be
affected through the integrated Sachs-Wolfe effect. Additionally,
there may be gravitational lensing effects. In general, the
quantitative effect of the inhomogeneities will depend on the
evolution of their amplitude and size. In our picture, the
amplitude of the inhomogeneities in $\rho_\Lambda$ is fixed.  The
size scale is determined by the number of nucleated bubbles $n_b$,
which in turn depends on the nucleation rate, and can be
calculated numerically \cite{ma05}. We shall carry out such
analysis elsewhere. Depending on the parameters of the model, a
wide range of sizes is possible. Hence, constraints on the range
of parameters may be obtained from compatibility with future
observations.

\section{The late-time phase transition} \label{models}

In order to compare the phase-coexistence scenario, in which the
hidden sector has a temperature $T=T_c\sim \rho_\Lambda^{1/4}$, to
the case in which $T_c\lesssim T\ll \rho_\Lambda^{1/4}$
\cite{chn04} and to that of supercooling, with $T\ll T_c\sim
\rho_\Lambda^{1/4}$ \cite{g00}, we need to consider a specific
model. Several models for late-time phase transitions have been
proposed in the literature (see
\cite{gc93,bs01,y02,g00,ahkm00,chn04,ch00} and references
therein). For our general considerations, a simple model in which
a single scalar field $\phi $ plays the role of an order parameter
will suffice.

\subsection{A potential with a negative mass squared}

The simplest potential we can consider is one with a negative
$\phi^2$ term, so the minimum is away from the origin;  thermal
effects then move the minimum back to the origin at high
temperature. Therefore, we consider the effective potential
\begin{equation}
V\left( \phi \right) =-\lambda v^{2}\phi ^{2}/2+\lambda \phi
^{4}/4+\rho_\Lambda ,  \label{pot}
\end{equation}
where the constant term $\rho_\Lambda \equiv \lambda v^{4}/4$ makes
the energy density vanish in the vacuum. This potential can be
regarded as a simplified version of the model considered in Ref.
\cite{chn04}, in which two scalar fields have negative squared
masses in the scale of $ TeV^{2}/M_{P} $ from supersymmetry
breaking.

The potential (\ref{pot}) does not even possess a metastable vacuum,
since the only minimum is $\phi =v$. However, at finite temperature,
the effective potential receives temperature-dependent corrections
from particles which have gauge or Yukawa couplings with $\phi $. We
will assume for simplicity that all the particles in the hidden
sector are bosons which couple to $\phi$ through field-dependent
masses $m\left( \phi \right) =h\phi $, all with the same coupling
$h$. The temperature-dependent part of the potential is (at 1-loop
order)
\begin{equation}
\Delta V_T (\phi)=\left( gT^4/2\pi^2\right) I\left(
h\phi/T\right),\label{fintemp}
\end{equation}%
where $I(x)=\int_0^\infty
dyy^2\log\left[1-\exp\left(-\sqrt{y^2+x^2}\right)\right]$, and $g$
is the  number of degrees of freedom in the hidden sector.

In the high temperature approximation [$m(\phi)\ll T$], the
function $I(x)$ is usually expanded in powers of $x$, and the free
energy of the hidden sector takes the form
\begin{equation}
V_{\textrm{high-}T}=\rho_\Lambda -g\pi ^{2}T^{4}/90+\Delta V\left(
\phi ,T\right), \label{vhight}
\end{equation}
where the field-dependent part
\begin{equation}
\Delta V\left( \phi ,T\right)=\frac{gh^{2}}{24}%
\left( T^{2}-T_{0}^{2}\right) \phi ^{2}-\frac{gh^{3}}{12\pi }T\phi
^{3}+\frac{\lambda }{4}\phi ^{4} \label{effpot2}
\end{equation}%
gives the free energy difference between the two phases. Here,
$T_{0}^{2}\simeq 12\lambda v^{2}/gh^{2}$. We have kept only the
relevant finite-$T$ contributions to $\Delta V$. One of them is of
the form $T^{2}\phi ^{2}$ and causes the vacuum expectation value
of $\phi $ to vanish at high $T$. The other one is of the form
$-T\phi ^{3}$ and produces a first-order phase transition. At the
critical temperature, $T_{c}=T_{0}/\sqrt{1-gh^{4}/6\pi ^{2}\lambda
},$ the high-temperature minimum $\phi=0$ becomes degenerate with
the low-temperature one, which at $T_c$ is given by
$\phi_{c}\equiv\left( gh^{3}/6\pi \lambda \right) T_{c} $. For
$T_{c}>T>T_{0}$ the two minima are separated by a barrier, which
at $T=T_{0}$ disappears, turning the minimum at $\phi =0$ into a
maximum.

For $T>T_{c}$, the hidden sector is in the phase with $\phi =0$.
In this case, $\Delta V=0$ in (\ref{vhight}) and, as expected, the
free energy gives an energy density of the form $\rho =\rho
_{\Lambda }+\rho _{R}$. Notice that, for ${\cal O}(1)$ couplings,
we have $T_{c}\sim T_{0}\sim v$, and $\rho_\Lambda \sim v^{4}$;
so, naturally, $T_{c}\sim \rho_\Lambda ^{1/4}\sim 10^{-3}eV$. If
we require instead that $T_{c}\ll 10^{-3}eV$, so that a vacuum
energy is attained through the condition $T_c<T<T_{\gamma }$
\cite{chn04}, then necessarily $T_{0}<\rho_\Lambda ^{1/4}/10$, and
we obtain the constraint $\lambda /g^{2}h^{4}<10^{-6}$, which is
not achieved with natural values of the parameters.

At high temperature the bubble nucleation rate is given by $
\Gamma \sim T^{4}\exp \left[ -F_c(T)/T\right] $, where $F_{c}$ is
the free energy of the critical bubble that is nucleated. $
F_{c}(T)$ diverges at $T=T_{c}$, and vanishes at $T=T_{0}$. Hence,
for $T$ close to $T_{c}$ the rate $\Gamma $ is exponentially
suppressed and there is always some supercooling. If the model
supports a large amount of supercooling, the hidden sector may
have $ \rho\simeq \rho _{\Lambda } $ at $T\ll T_{c}$. Notice
however that, for $T\simeq T_{0}$ the nucleation rate is extremely
high, $\Gamma \sim T^{4}$, so the phase transition will  certainly
end before $T$ gets close to $T_{0}$. Indeed, we may roughly
consider that bubble nucleation effectively begins when its rate
becomes $\Gamma \sim H^{4}$. Using $H^2=8\pi \rho /3M_{P}^{2}$,
with $\rho \sim 10^{3}\rho _{\Lambda }$ for $T\sim 10^{-3}eV$, we
see that the onset of nucleation happens as soon as $F_c/T$
becomes smaller than $\simeq 270$.

If we require supercooling at $T<T_{\gamma }\sim 10^{-4}eV$, then
we must have, on one hand, $T_{0}< T< T_{c}/10$.  This condition
will be unnatural in any model, since in general $T_0\sim T_c\sim
\rho_\Lambda ^{1/4}$. As we have seen, the condition
$T_0<\rho_\Lambda ^{1/4}/10$ already constrains the parameters. If
we require also $T_0<T_c/10$ we find for our model the further
constraint $1-10^{-2}<gh^4/6\pi^2\lambda< 1$, so the parameters
must be fine tuned with a precision of $10^{-2}$. This constraint
is not compatible with the previous one. Of course, this is due to
the simplicity of the model we are considering. Below, we consider
a model in which $T_c$ and $T_0$ are independent of each other.
Notice, however, that even disregarding one of the two
constraints, the other one still implies either unnatural values
of the parameters or a fine tuning.

Besides, we must demand that $\Gamma <H^{4}$ still at $T<
10^{-4}eV$. With $\rho \sim \rho _{\Lambda }$, this requires
$F_c/T> 270$, which is quite a large value for $F_c$ if we take
into account that $T\ll T_c$. The free energy $F_c$ is difficult
to estimate analytically. In the thin wall approximation, it is
given by $ F_c(T)/T\simeq \left( 16\pi /3\right) \sigma
^{3}T_{c}/L^{2}\left( T_{c}-T\right) ^{2}$, where $\sigma$ is the
bubble wall tension. Although $F_c\to \infty$ for $T\to T_{c}$, in
the case $T\ll T_{c}$, a large enough value of this quantity can
only be attained with an unnaturally large value of the surface
tension $\sigma $ \cite{g00}. In fact, the thin wall approximation
is no longer valid in this limit, and a numerical calculation is
required. For a simple potential of the form of Eq.
(\ref{effpot2}), such calculation has been performed
\cite{dlhll92, m00}. In the case $T_0\ll T_c$, the free energy of
the critical bubble is given by
\begin{equation}
F_c(T)/T=7.7 \left(
g^{1/4}/\lambda^{3/4}\right)\alpha^{3/2}f(\alpha), \label{linde}
\end{equation}
where $\alpha=\left(T^2-T_0^2\right)/T^2$. In Eq. (\ref{linde}) we
have used the constraint $gh^4/6\pi^2\lambda\simeq 1$ to eliminate
the parameter $h$. In Ref. \cite{dlhll92}, it was found that the
function $f(\alpha)$ is fit by the expression
\[
f(\alpha)\simeq 1+\frac{\alpha}{4}\left(1+\frac{2.4}{1-\alpha}+
\frac{0.26}{(1-\alpha)^2}\right)
\]
with an accuracy of 2\%. Assume for definiteness that $g\sim 10$.
Then, from Eq. (\ref{linde}) it is not difficult to find the
constraints imposed by the condition $F_c/T> 270$. On one hand, if
$\lambda={\cal O}(1)$, such high value of $F_c/T$ is attained only
for $T>10T_0$, so $T_0$ must be less than $10^{-2}T_c$. This
increases the fine tuning on the parameters. On the other hand,
for $T\sim T_0$ the condition is reached for unnaturally small
values of the parameter $\lambda$ (e.g., $T\simeq 1.5T_0$ requires
$\lambda <10^{-2}$).

Notice that immediately after the completion of the phase
transition there is still a non-vanishing vacuum energy. Indeed,
the value $\phi_c=\phi(T_c)$  of the free energy minimum at the
critical temperature is displaced from the zero-temperature (true
vacuum) value $\phi(T=0)=v$. However, after the phase transition
$\phi (T)$ evolves with temperature and the energy density $\rho
_{\mathrm{vac}}(\phi)$ does not behave like a cosmological
constant. During the phase coexistence period, instead, as one
phase is converted into the other, the values $\phi=0$ and
$\phi=\phi_c$ coexist at $T_c$, and one expects that the average
vacuum energy remains larger than a certain value $\rho
_{\mathrm{vac}}=\mathcal{O}\left( \rho_\Lambda \right) $. Indeed,
since $\rho_\Lambda \sim \lambda v^{4}$ and $T_{c}^{4}\sim
T_{0}^{4}\simeq \left( 12\lambda /g\right) ^{2}v^{4}/h^{4}$, Eq.
(\ref{Leff}) gives a positive $\rho_{\Lambda }^\textrm{eff}\sim
\lambda v^{4}$ as long as the coupling $h$ is not too small (i.e.,
if $h\simeq 1$).

According to Eq. (\ref{lrho}), the period of phase coexistence can
be long enough if the latent heat is close to its maximum value
$L_\textrm{max}\sim \rho_R$. The latent heat $L\equiv T\partial
\Delta V/\partial T|_{T=T_{c}}$ is readily calculated for the
potential (\ref{effpot2}). For $ T_{c}\sim T_{0}$ and $\phi
_{c}\sim v$, we have $L \simeq \lambda v^{2}\phi _{c}^{2}$, so in
order to accomplish $L\sim\rho _{R}$ we obtain the condition $\phi
_{c}/T_{c}\gtrsim 1$, i.e., the phase transition must be strongly
first-order. This is achieved for strong couplings $h\gtrsim 1$.
Notice that within this approximation the latent heat $L$ is
unbound. In fact, for $h\phi/T>1$ the high-temperature
approximation (\ref{effpot2}) breaks down. If we use the exact
one-loop result (\ref{fintemp}), the latent heat is given by
\begin{equation}
L=\frac{2gT^4}{\pi^2}\left[-
 I\left(0\right)+I\left(\frac{h\phi_c}{T_c}\right)-
\frac{h\phi_c}{4T_c}
I^{\prime}\left(\frac{h\phi_c}{T_c}\right)\right] .
\end{equation}%
The functions $I(x)$ and $I^{\prime}(x)$ decay exponentially for
$x$ large. Hence, we can get arbitrarily close to the maximum
value $L_{\max}=-2I(0)gT^4/\pi^2$ by increasing  $x$ [e.g., for
$x=10$ we have  $I(10)/I(0)\sim 10^{-3}$]. A high value of
$x\equiv h\phi_c/T_c$ is again achieved with a strong coupling
$h$, since the order parameter $\phi_c/T_c$ increases with $h$.
Perturbativity imposes a generic upper limit $h\lesssim
\sqrt{4\pi}$. However, this limit depends on the details of the
model \cite{cmqw05}.

\subsection{A potential with Coleman-Weinberg symmetry breaking}

Alternatively, one can obtain a strong phase transition by
considering a gauge theory with a strong coupling scale $\sim
\rho_{\Lambda}^{1/4}$. A model in which a hidden $SU\left(
2\right) $ Yang-Mills theory has a chiral phase transition at a
scale $\Lambda _{SU\left( 2\right) }\sim 10^{-3}eV$ has been
considered by Goldberg \cite{g00}. The model has $g=34$  d.o.f.,
comprising the $SU\left( 2\right) $ gauge fields and 8 Weyl
doublets. In Ref.
\cite{g00} it is shown that in this case the supercooling condition is $%
T/\rho _{\Lambda }^{1/4}\lesssim 10^{-2}$.

A linear sigma model gives an effective potential of the form
\begin{equation}
V\left( \phi ,T\right) =A\left( T^{2}-T_{0}^{2}\right) \phi
^{2}/2+\lambda \phi ^{4}\left[ \log \left( \phi /\phi _{0}\right)
-1/4\right] , \label{vgold}
\end{equation}%
where $A$ depends on the number of fermions; in this case, $A\simeq
24\sqrt{\lambda }$ (see \cite{g00} for details). Notice that the
temperature-dependent term here corresponds to the first term in the
high-temperature expansion (\ref{effpot2}). Furthermore, the role of
$T_{0}$ in the dynamics of the phase transition is similar to the
one it played in the previous model: there is a barrier between the
two vacua for $T_{0}<T<T_{c}$. So, in order to achieve the desired
supercooling, the condition $T_{0}\ll T_{c}$ must be imposed. There
is no reason for such a hierarchy, and in general it will be
difficult to realize in a realistic model \cite{g00}. Nevertheless,
in Eq. (\ref{vgold}) the parameter $T_{0}$ can be set to $0$, since
the symmetry is already broken spontaneously in the Coleman-Weinberg
manner by the last term. Setting $T_{0}=0$, the vacuum energy is
given by the difference $V(0)-V(\phi_0)$ at $T=0$. It is related to
$T_c$ by $\rho_\Lambda=(4.4T_c)^4$. It can be then shown that the
required supercooling is attained with $\lambda \leq 0.01$.

On the other hand, the latent heat is easily calculated for the free
energy (\ref{vgold}); we obtain $L=AT_{c}^{2}\phi _{c}^{2}$. Here,
the critical temperature and order parameter are related by $\phi
_{c}^{2}=A\left( T_{c}^{2}-T_{0}^{2}\right) /\lambda $. Setting
$T_{0}=0$, we find $L=A^{2}T_{c}^{4}/\lambda \simeq 560T_{c}^{4}$,
which gives the ratio $L/\rho _{R}\simeq 50$. Obviously, this value
of $L$ is unrealistically large. This means that the
high-temperature approximation is not valid, so expression
(\ref{fintemp}) should be used for the finite-$T$ correction. This
fact indicates that supercooling demands an extremely strong phase
transition.

Since the supercooling requirement implies such a large $L$, we
infer that  the conditions for phase coexistence will be less
stringent in this model. If supercooling is not needed, the
restrictions $T_{0}\ll T_{c}$, $\lambda \ll 1$ can be relaxed.
Without entering into the details of the model, we can  consider
the condition $L/\rho _{R}\sim 1$ to get an idea of the
requirements of a long enough phase-coexistence stage. Letting $
T_{0}\neq 0$, we have the condition $ 50\left(
T_{c}^{2}-T_{0}^{2}\right) /T_{c}^{2}\sim 1$, which is independent
of $\lambda $ and is satisfied even for $T_{0}$ very close to
$T_{c}$. This shows that the phase coexistence scenario arises
naturally in this model.

The comparison between the supercooling and phase-coexistence
scenarios is transparent in this case because the latent heat is
proportional to the difference $T_{c}^{2}-T_{0}^{2}$, which is
required to be unnaturally large in the supercooling scenario.
Thus, a large amount of supercooling is clearly more difficult to
attain than a sizeable latent heat, and the phase coexistence
scenario is favored. However, to get a precise comparison of the
two scenarios a numerical calculation of the phase transition
would be suitable. We shall perform such calculation elsewhere. In
any case, both supercooling and phase-coexistence may occur in
general, and both effects contribute to delay the completion of
the phase transition.

\section{Conclusions} \label{conclu}

We have pointed out that a late-time first-order phase transition
may enter a long phase-equilibrium stage, thus avoiding the need
of excessive supercooling to account for the durability of a false
vacuum phase. Basically, this occurs if the cooling method used by
the universe (namely, the adiabatic expansion of a relativistic
gas) fails in taking away the latent heat associated to the
transition. Thus, the main requirement for this scenario is that
the latent heat must be of the order of the energy density of
radiation. We have shown that if the temperature scale of the
phase transition is $T_{c}\sim 10^{-3}eV$, the current equation of
state for phase coexistence is essentially $\omega=-1$, and the
dark energy density has the correct magnitude. The distinctive
inhomogeneities in $\rho_\textrm{vac}$ during phase coexistence
may leave an imprint on large scale structure. We have argued that
a long period of phase coexistence may arise naturally in the
context of a particle physics theory, in contrast to the
supercooling case. In general, the condition $L\sim \rho _{R}$ is
achieved in theories with a strong coupling $h\gtrsim 1$.

\section*{Acknowledgements}

I thank C. Biggio, E. Mass\'o, G. Zsembinszki, and especially J.
Garriga for helpful conversations. I also thank D. Scott for useful
comments.

\end{document}